\documentclass[english,a4paper,eqno,11pt]{article}
\usepackage[english]{babel}
\usepackage{graphicx,epsfig,epic} 
\usepackage{pifont}      

\usepackage{pict2e}
\usepackage{amssymb}

\font\tenbb=msbm10 at 12pt
\def\cC{\hbox{\tenbb C}}
\def\rR{\hbox{\tenbb R}}

\def\cal{\mathcal}
\font\titre cmbx10 at 18 pt 
 
\def\esp{\vskip .6cm}
\def\pesp{\vskip .3cm}

\newtheorem{rem}{Remark}

\newsavebox{\fmbox}

\begin{document}
\centerline{\titre {Space-Time Geodesics and the Derivation}}
\pesp
\centerline{\titre {of Schr\"{o}dinger's equation}}
\pesp

\esp
\centerline{\small Fay\c cal BEN ADDA}
\pesp
\centerline{\small \it Community College of Qatar}
\centerline{\small\it  f$\_$benaddafr@yahoo.fr}
\pesp
\centerline{(\today)}

\esp

{\small \bf Abstract}
Using the essence of Feynman's path integral and the space-time geodesics, an infinity of differentiable paths that follow the geometry of a continuous geodesic are constructed, and a wave function is associated to each path as a probability amplitude identical to the Feynman's probability amplitude for each path. We prove that each probability amplitude obeys to the Schr\"{o}dinger's equation for a non relativistic physical system moving in a time independent potential, starting from the Jacobi-Hamilton equation.

\pesp


\esp
{\small\tableofcontents}
\section{Introduction}
The most challenging problem of modern physics is to find a coherent geometrical representation of the space-time at all scales. The real geometry and architecture of the space-time of our universe and the constraints that it imposes to its content are still unknown despite the fact that local descriptions of the space-time geometry provide a suitable understanding of the behavior of several physical systems on it; the space-time is well described by an Euclidian space-time with null curvature between approximatively $10^{-18}m$ and $10^{11}m$, meanwhile the same space-time becomes well described by a continuous pseudo-Riemannian manifold with non null curvature for the scales between approximatively $10^{11}m$ and $10^{25}m$. The two local descriptions of the same space-time are deeply different, meanwhile the continuity of the medium seems to be the common factor. The  study of curvature, evolution, characteristics and expansion of the different geometrical representations of our space-time allows to acquire a coherent understanding of the local representations of the space-time architecture and geometry, however it barely provides a global understanding of the space-time geometry and architecture. \pesp

Without knowing the real constraints that the space-time geometry and characteristics impose to its content, we may not be able to acquire a precise understanding of certain physical systems behaviors. An approximation based on our satisfactory local models for a specific range of scales would be considered suitable, until new leads may transcend  our understanding of the physical phenomena. One of these new leads is the Feynman's description of the quantum evolution that creates a bridge between the Schr\"{o}dinger's formulation of quantum mechanics and classical mechanics.
\pesp

Heisenberg, Born and Jordan (\cite{BHJ},\cite{BJ},\cite{Hei}) have developed a matrix formalism of quantum mechanics in 1925 in which matrix was used as a physical property of elementary particles. In 1926 Schr\"{o}dinger introduced another approach of quantum mechanics, where the physical properties of elementary particles were described by a wave function solution of a differential equation (\cite{SCH1,SCH2,SCH3,SCH4}). The Schr\"{o}dinger's approach found the development of its formalism on an infinitely dimensional Hilbert space of functions, and since the matrix representation of the quantum physical properties was recognized as a representation of the wave function in a Hilbert space, the two approaches of quantum mechanics were unified in the Hilbert space formalism. The quantum description of a point particle is completely unusual for a trajectory basic understanding; indeed since the quantum formalism prohibits the simultaneous determination of position and momentum with accurate measurement, therefore the classical trajectory of a particle has no physical meaning in the quantum mechanic's interpretation. \pesp

In 1948 Feynman introduced a different approach of quantum mechanics using the path integral (\cite{FEY1,FEY2}) as a sum over all possible trajectories to calculate a total quantum amplitude compatible with the Schr\"{o}dinger's equation. Feynman's path integral conveys a connection between Lagrangian and quantum mechanics that can be traced back to 1933 in an earlier work of Dirac \cite{DI}, in which the Huygens' principle for the interpretation of the wave nature of light was mathematically formulated in an integral equation of the form
\begin{equation}
\Psi(x,t_1)=\int G(x,y)\Psi(y,t_0)dy,\qquad\hbox{with}\quad t_1>t_0,
\end{equation}
where $G(x,y)$ was called the propagator (or kernel), and for which Dirac suggested that the propagator was analogous to the complex function $e^{{iS\over \hbar}}$. Feynman \cite{FEY3} found that the propagator was rather approximatively equal to $e^{{iS\over \hbar}}$, and found out the way to connect approximatively the Lagrangian to the Schr\"{o}dinger's formulation of quantum mechanics. Nevertheless, despite the correct multiple predictions of the Feynman's path integral in physics, a lack of mathematical rigor exists for the geometrical interpretation of the integration over all possible paths.\pesp

The main objective of this work is to use the essence of the Feynman's path integral in a different way that involves the medium characteristics to provide a fully equivalent meaning compatible with the Schr\"{o}dinger's formulation of quantum mechanics. The most appealing characteristics of Feynman's path integral are the use of an infinity of classical trajectories, the association of a probability amplitude to each trajectory via the classical action function, and the summation of the infinity of probability amplitudes over all possible paths to represent the total probability amplitude of finding a physical system in a certain location compatible with the Schr\"{o}dinger's equation. It sounds quiet simple but deeply difficult to assemble with mathematical rigor, and different attempts to give meaning to Feynman's approach are still under intense research and development. Some approaches can be found in the following list but not limited to \cite{AW5},\cite{BA5},\cite{FH5},\cite{RH5},\cite{Mk},\cite{KLMP5},\cite{kro5},\cite{nel5},\cite{SO5}.\pesp

The main difficulty in any approach to the Feynman's path integral is the infinity of all possible trajectories. They are classical trajectories but unknown, with no connection between them or continuous transformations that allow to pass from one path to another to buildup an adequate metric for any integration over all of them; even to define all trajectories via a given action function remains insufficient to reflect the space-time print. This particular difficulty will be circumvented and presented in a different and more accessible way. For example, if the space-time is defined by the surface of a right circular cylinder, all possible paths on the cylinder to join two different locations on the cylinder obey to the constraints that the geometry of the surface of the cylinder imposes to them; the print of the cylinder is given by its geodesics which are straight line segments parallel to the centerline of the cylinder and arcs of circle perpendicular to the centerline. Hence any possible path on the cylinder must be expressed by a combination of the two geodesics of the cylinder. Then in general if all possible trajectories must exist in the same space-time representation, then the print of the space-time (the constraints that the geometry and characteristics of the space-time impose) must be the common connection between all of them. This connection should be conveyed in the formalism that involves all the possible paths that might be followed by a given physical system. \pesp

 For this purpose, a case study of an infinity of classical trajectories between initial and final states of a given physical system is expressed in term of graphs of differentiable functions that follow the geometry of the graph of a single continuous geodesic; where the continuous function represents the print of the space-time that connects all possible trajectories (the space-time geodesic that models all possible paths). An example of modeling an infinity of differentiable paths that follow the geometry of a given continuous and nowhere differentiable function can be found in \cite{BFP}. \pesp

Since the common factor between our suitable local representations of the space-time is the continuity of the medium, then the continuity will be taken as the general characteristic of the space-time print (the space-time geodesics), and it is represented in this case study by a continuous function that models all possible paths by its geometry (modeling all possible paths by the geometry of a continuous function is described within this work by the intermediate of integral functions).
Nevertheless this continuous function can be replaced by any other continuous function combination of space-time geodesics that reflects a more complex space-time print at any range of scales with more specific constraints such as regularity, fluctuation, oscillation, smoothness by intervals, nowhere differentiability, variable curvature, etc.
\pesp

The plan of this paper is as follow: in section 2, an infinity of differentiable paths is constructed that follow the geometry of a given continuous function. In section 3, to each path a wave function is associated as a probability amplitude identical to the Feynman's probability amplitude for each path. The compatibility with the indetermination principle is discussed in section 4. In section 5, one prove that each probability amplitude obeys to the Schr\"{o}dinger's equation for a non relativistic physical system moving in a time independent potential, starting from the Jacobi-Hamilton equation. A general prospect is given in section 6.

\section{Infinity of Paths Related to the Geometry of a Continuous Function}\label{subect0}

\pesp

Let us consider a continuous function
\begin{eqnarray}\label{Ndp}
   x:I& \rightarrow &\rR\nonumber \\
        t &\mapsto & x(t),
\end{eqnarray}
 where $I$ an open interval of $\rR$. Assuming that the graph of the function (\ref{Ndp}) for all $t\in[t_a,t_b]\subset I$ represents the local print of a given space-time representation (the space-time geodesics), where $t_a,t_b\in\rR$ such that $x(t_a)=a$ and $x(t_b)=b$ for a given pair of real numbers $a$ and $b$.\pesp

Consider a physical system of mass $m$ moving in such a space-time with the print function given by (\ref{Ndp}) on $[t_a,t_b]$, and suppose that the motion of the physical system is completely unknown. Consider an infinity of possible trajectories defined in the space-time that follow the continuous function (\ref{Ndp}) on $[t_a,t_b]$. These infinity of trajectories can be defined by the graph of the function
\begin{equation}\label{xd}
x_\delta(t):=x(t,\delta)={1\over\delta}\int_t^{t+\delta}x(t') dt'
 \end{equation}
 that verifies
\begin{equation}\label{F0}
x_0(t)=\lim_{\delta\rightarrow 0}x_\delta(t)=x(t),
\end{equation}
for all $t\in\ [t_a,t_b]$ and for all $\delta\in\ ]0,\alpha]$, where  $\alpha>0$ is an arbitrary small real number.\pesp

Based on the definition (\ref{xd}), the print function (\ref{Ndp}) can be formulated in terms of an infinity of paths defined by the graph of the functions (\ref{xd})
since we have for all $t\in\ [t_a,t_b]$ and for all $\delta\in\ ]0,\alpha]$
\begin{equation}\label{xt}
x(t)=x_\delta(t)+\delta\Big(\frac{\partial x_\delta(t)}{\partial \delta}- \frac{\partial x_\delta(t)}{\partial t}\Big),
\end{equation}
or it can be written in terms of operator applied to the function $x(t,\delta)$ as
\begin{equation}\label{xt1}
x(t)=\Big(Id+\delta(\frac{\partial }{\partial \delta}- \frac{\partial}{\partial t})\Big)x(t,\delta),
\end{equation}
where $(Id) x(t,\delta)=x(t,\delta)$ and $\delta(\frac{\partial }{\partial \delta}- \frac{\partial}{\partial t})x(t,\delta)=\delta\Big(\frac{\partial x(t,\delta)}{\partial \delta}- \frac{\partial x(t,\delta)}{\partial t}\Big)$.

Meanwhile for one trajectory given by (\ref{xd}) for $\delta=\delta_0$

\begin{equation}
x(t)\neq x_{\delta_0}(t)+\delta_0\Big(\frac{\partial x_{\delta_0}(t)}{\partial \delta}- \frac{\partial x_{\delta_0}(t)}{\partial t}\Big),
\end{equation}
that is to say the continuous function (\ref{Ndp}) cannot be expressed in term of one smooth trajectory given by $x_{\delta_0}(t)$ for all $[t_a,t_b]$. The use of an infinite of trajectories is defined implicitly in the equality (\ref{xt}).\pesp

The velocity of the physical system that follows all possible trajectories given by (\ref{xd}) for each $\delta\in\ ]0,\alpha]$ and for all $t\in\ [t_a,t_b]$ is given by the rate of change of the space-time print (\ref{Ndp})
\begin{equation}\label{Vd}
v_{\delta}(t):={\partial  x_{\delta}(t)\over\partial t}={x(t+\delta)-x(t)\over \delta}.
\end{equation}
The process of defining an infinity of possible trajectories in a given space-time that follows the geometry of a continuous function (\ref{Ndp}) can be repeated to obtain an infinity of  ${\cal C}^2$ functions that follows the geometry of (\ref{Ndp}):\pesp

Consider for all $\delta\in ]0,\alpha]$ and for all $\varepsilon\in]0,\beta]$, where $\beta>0$ is an arbitrary small real number, the function
$x_{\delta\varepsilon}(t):=x(t,\delta,\varepsilon)$ defined by
\begin{eqnarray}\label{xde}
   x_{\delta\varepsilon}:[t_a,t_b]& \rightarrow &\rR\nonumber \\
        t &\mapsto & x_{\delta\varepsilon}(t)={1\over\varepsilon}\int_t^{t+\varepsilon}x_{\delta}(t') dt'.
\end{eqnarray}

A direct calculus for all $t\in\ [t_a,t_b]$, for all $\delta\in\ ]0,\alpha]$, and for all $\varepsilon\in\ ]0,\beta]$, gives
\begin{equation}
x=x_{\delta\varepsilon}(t)+\varepsilon\Big(\frac{\partial x_{\delta\varepsilon}(t)}{\partial \varepsilon}- \frac{\partial x_{\delta\varepsilon}(t)}{\partial t}\Big)+\delta\Big(\frac{\partial x_\delta(t)}{\partial \delta}- \frac{\partial x_\delta(t)}{\partial t}\Big).
\end{equation}
that is to say the space-time print function can be formulated in terms of an infinity of paths defined by the graph of the functions (\ref{xde})  for all $t\in[t_a,t_b]$, for all $\delta\in\ ]0,\alpha]$, and for all $\varepsilon\in\ ]0,\beta]$.\pesp

Suppose that all possible trajectories, for a given physical system with mass $m$ moving in the space-time representation with the continuous print function (\ref{Ndp}), are defined by the infinity of trajectories given by the graph of the continuous and differentiable functions (\ref{xde}) for all $t\in\ [t_a,t_b]$, for all $\delta\in ]0,\alpha]$ and for all $\varepsilon\in]0,\beta]$. The velocity ${\cal V}_{\delta\varepsilon}(t)$ of the physical system  for all $\delta\in\ ]0,\alpha]$, and for all $\varepsilon\in\ ]0,\beta]$ is then defined by
\begin{eqnarray}\label{Vdef}
   {\cal V}_{\delta\varepsilon}:[t_a,t_b]& \rightarrow &\rR\nonumber \\
        t &\mapsto & {\cal V}_{\delta\varepsilon}(t)={\partial x_{\delta\varepsilon}(t)\over\partial t},
\end{eqnarray}
then
\begin{equation}\label{Vde}
{\cal V}_{\delta\varepsilon}(t)= {x_{\delta}(t+\varepsilon)-x_{\delta}(t)\over\varepsilon}.
\end{equation}
\pesp
Despite the fact that the position (\ref{xde}) and the velocity (\ref{Vde}) are known at any time, the indeterminacy resides in the location at a given scale $\delta$ and a given scale $\varepsilon$ among the infinity of scales $\delta\in\ ]0,\alpha]$ and the infinity of scales $\varepsilon\in\ ]0,\beta]$, which is consistent with the indetermination principle of quantum mechanics. It will be discussed in Section 4.

\section{Wave Function Associated to the Infinity of Paths}

The consideration of an infinity of possible paths for the quantum physical system leads to associate to each path a probability amplitude (following the essence of Feynman's path integral) that corresponds to the probability of finding the physical system in a certain location at a given scale. For this purpose we introduce the following: \pesp

Let $\Phi$ be the potential function defined for all $\delta\in\ ]0,\alpha]$, and for all $\varepsilon\in\ ]0,\beta]$ by
\begin{eqnarray}\label{Pot}
   \Phi:\rR\times[t_a,t_b]& \rightarrow &\rR\nonumber \\
        (x_{\delta\varepsilon}(t),t) &\mapsto & \Phi(x_{\delta\varepsilon}(t),t).
\end{eqnarray}

Let us consider for all $\delta\in\ ]0,\alpha]$, and for all $\varepsilon\in\ ]0,\beta]$ the action function
\begin{eqnarray}\label{Act}
   {\cal S}:\rR\times [t_a,t_b]& \rightarrow &\rR\nonumber \\
        (x_{\delta\varepsilon}(t),t) &\mapsto & {\cal S}(x_{\delta\varepsilon}(t),t),
\end{eqnarray}
related to the momentum ${\cal P}$ and total energy $E$ for all $t\in[t_a,t_b]$, for all $\delta\in\ ]0,\alpha]$, and for all $\varepsilon\in\ ]0,\beta]$ by the following equations
\begin{eqnarray}
  {\partial{\cal S}\over\partial x_{\delta\varepsilon}} &=&m{\cal V}_{\delta\varepsilon}= {\cal P}\label{4Act1} \\
  -{\partial{\cal S}\over\partial t} &=& E. \label{4Act2}
\end{eqnarray}

Let us consider for all $\delta\in\ ]0,\alpha]$, and for all $\varepsilon\in\ ]0,\beta]$ the following wave function associated to each path defined by (\ref{xde})
\begin{eqnarray}\label{W}
   \Psi:\rR\times [t_a,t_b]& \rightarrow &\cC\nonumber \\
        (x_{\delta\varepsilon}(t),t) &\mapsto & \Psi(x_{\delta\varepsilon}(t),t)=e^{{i{\cal S}(x_{\delta\varepsilon}(t),t)\over \hbar}},
\end{eqnarray}
where the symbol $i$ is the imaginary unit, $\hbar={h\over2\pi}$ and $h$ is the Plank's constant. The function $\Psi(x_{\delta\varepsilon},t)$ represents the wave function for a single particle where $x_{\delta\varepsilon}(t)$ is its position in space at the time $t\in[t_a,t_b]$, the scale $\delta\in\ ]0,\alpha]$, and the scale $\varepsilon\in\ ]0,\beta]$ (this wave function was used by Feynman in his postulate II of the path integral). Then similarly to Feynman's choice, the  wave function (\ref{W}) represents a probability amplitude that corresponds to the probability of finding the physical system in a certain location at the time $t$, the scale $\delta\in\ ]0,\alpha]$, and the scale $\varepsilon\in\ ]0,\beta]$.
The wave function $\Psi(x_{\delta\varepsilon},t)$ may take the approximate form
\begin{equation}
\Psi(x_{\delta\varepsilon},t)=Ae^{{i{\cal S}(x_{\delta\varepsilon},t)\over \hbar}}
\end{equation}
where $A$ is the dimensional constant required for any physical interpretation. In the following the amplitude $A$ of the used wave function will be considered as equal to one for more simplicity and it can be reintroduced if needed in any linear equation.\pesp

\begin{rem}
The continuous function (\ref{Ndp}) is postulated as the geodesic of the considered space-time (the space-time print), that is to say in the Lagrangian approach the continuous function (\ref{Ndp}) is an extremum of an action function  $S(x(t),t)$. Then every small deviation from the path of (\ref{Ndp}), such as the infinity of paths (\ref{xde}) for all $\delta\in\ ]0,\alpha]$, and all $\varepsilon\in\ ]0,\beta]$, will result in a larger (or smaller) action function (\ref{Act}).
\end{rem}
The wave function (\ref{W}) gives, using equations (\ref{4Act1}) and (\ref{4Act2}), the correspondence principles of quantum mechanics for momentum and total energy
for all $t\in[t_a,t_b]$, for all $\delta\in\ ]0,\alpha]$, and for all $\varepsilon\in\ ]0,\beta]$
\begin{equation}\label{Cp}
{\cal P}=-i\hbar {\partial\Psi(x_{\delta\varepsilon}(t),t)\over\partial x_{\delta\varepsilon}}{1\over \Psi(x_{\delta\varepsilon}(t),t)},
\end{equation}
and
\begin{equation}\label{Ce}
E=i\hbar {\partial\Psi(x_{\delta\varepsilon}(t),t)\over\partial t}{1\over \Psi(x_{\delta\varepsilon}(t),t)}.
\end{equation}

\section{Compatibility with the Indetermination Principle}\label{Compat}

The classical trajectories are used in a way which remains compatible with the indetermination principle; this can be seen in the computation of the probability amplitude. Indeed, the wave function $\Psi(x_{\delta\varepsilon},t)$ is interpreted as a probability amplitude that represents the wave function for a single particle, in which $x_{\delta\varepsilon}(t)$ is its position in space at the time $t\in[t_a,t_b]$, the scale $\delta\in\ ]0,\alpha]$ and the scale $\varepsilon\in\ ]0,\beta]$, and where the square modulus of the wave function
\begin{equation}\label{Mod}
\vert\Psi(x_{\delta\varepsilon},t)\vert^2=\Psi(x_{\delta\varepsilon},t)\Psi^*(x_{\delta\varepsilon},t)=\rho(x_{\delta\varepsilon},t),
 \end{equation}
for $\Psi^*$ the conjugate of $\Psi$, represents the probability density that the physical system at the time $t\in[t_a,t_b]$, the scale $\delta\in\ ]0,\alpha]$ and the scale $\varepsilon\in\ ]0,\beta]$ is located at $x_{\delta\varepsilon}(t)$.\pesp

If one measures the position of the physical system at the time $t\in[t_a,t_b]$, the scale $\delta\in\ ]0,\alpha]$ and the scale $\varepsilon\in\ ]0,\beta]$, then the location of the physical system is expressed by the probability distribution:\pesp

The probability that the physical system's position $x_{\delta\varepsilon}(t)$ is between $x_{\delta\varepsilon_0}(t)=c$ and $x_{\delta\varepsilon_1}(t)=d$ is given by the following integral of density
\begin{equation}\label{Prob01}
P^{\delta}_{c\leq x_{\delta\varepsilon}\leq d}(t)=\int^d_c \vert\Psi(x_{\delta\varepsilon},t)\vert^2 dx_{\delta\varepsilon}=\int^{\varepsilon_1}_{\varepsilon_0} \vert\Psi(x_{\delta\varepsilon},t)\vert^2 {\partial x_{\delta\varepsilon}(t)\over\partial \varepsilon}d\varepsilon,
\end{equation}
where $t$ and $\delta$ are the time and the scale at which the physical system is measured.\pesp
The normalization condition is then
\begin{equation}
\int^{+\infty}_{-\infty} \vert\Psi(x_{\delta\varepsilon},t)\vert^2 dx_{\delta\varepsilon}=1.
\end{equation}

The probability of distribution of location that depends on the variation of $\varepsilon$ is sufficient for the derivation of Schr\"{o}dinger's equation from Hamilton-Jacobi equation, and the indeterminacy resides in the location at a given scale $\varepsilon$ among the infinity of scales $\varepsilon\in\ ]0,\beta]$, which is consistent with the indetermination principle of quantum mechanics.
\pesp
\begin{rem} The probability of distribution of other possible locations can be summarized in the following but are not used in the development of this formalism:

i) The probability that the physical system's position $x_{\delta\varepsilon}(t)$ is between $x_{\delta_0\varepsilon}(t)=a'$ and $x_{\delta_1\varepsilon}(t)=b'$ is given by the following integral of density
\begin{equation}\label{Prob2}
P^\varepsilon_{{a'\leq x_{\delta\varepsilon}\leq b'}}(t)=\int^{b'}_{a'} \vert\Psi(x_{\delta\varepsilon},t)\vert^2 dx_{\delta\varepsilon}=\int^{\delta_1}_{\delta_0} \vert\Psi(x_{\delta\varepsilon},t)\vert^2 {\partial x_{\delta\varepsilon}(t)\over\partial \delta}d\delta,
\end{equation}
where $t$ and $\varepsilon$ are the time and the scale at which the physical system is measured.\pesp
The normalization condition is then
\begin{equation}
\int^{+\infty}_{-\infty} \vert\Psi(x_{\delta\varepsilon},t)\vert^2 dx_{\delta\varepsilon}=1.
\end{equation}

ii) The probability that the physical system's position $x_{\delta\varepsilon}(t)$ is between $x_{\delta_0\varepsilon_0}(t)=A$ and $x_{\delta_1\varepsilon_1}(t)=B$ is then given by the following integral of density
\begin{equation}
P_{{A\leq x_{\delta\varepsilon}\leq B}}(t)=\int^B_A \vert\Psi(x_{\delta\varepsilon},t)\vert^2 dx_{\delta\varepsilon}=P^\delta_{[a,b]}(t)+P^\varepsilon_{[a',b']}(t),
\end{equation}
 where $t$ is the time, $\delta$ and $\varepsilon$ are the scales at which the physical system is measured, and where $a,\ b,\  a',\ b'$ are defined above.\pesp

The normalization condition is then
\begin{equation}
\int^{+\infty}_{-\infty} \vert\Psi(x_{\delta\varepsilon},t)\vert^2 dx_{\delta\varepsilon}=1.
\end{equation}

iii) The probability that the physical system's position $x_{\delta\varepsilon}(t)$ is between $x_{\delta\varepsilon}(t_1)=a_1$ and $x_{\delta\varepsilon}(t_2)=a_2$ is then given by the following integral of density
\begin{equation}
P_{{a_1\leq x_{\delta\varepsilon}\leq a_2}}(t)=\int^{a_2}_{a_1} \vert\Psi(x_{\delta\varepsilon},t)\vert^2 dx_{\delta\varepsilon},
\end{equation}
 where $t$ is the time, $\delta$ and $\varepsilon$ are the scales at which the physical system is measured. The normalization condition is then
\begin{equation}
\int^{+\infty}_{-\infty} \vert\Psi(x_{\delta\varepsilon},t)\vert^2 dx_{\delta\varepsilon}=1,
\end{equation}
in this case there isn't any measure over variation of scales.
\end{rem}

\section{Derivation of Schr\"{o}dinger's Equation for a Non-relativistic Physical System Moving in a Time Independent Potential}\label{JHE}

Based on (\ref{Prob01}) the scale $\delta$ will be limited to a given small value, while the infinity of paths are  expressed in term of infinity of scales $\varepsilon$ in $]0,\beta]$, then in the following we consider only the case where the wave function $\Psi$ is defined by (\ref{W}) for a given $\delta\in\ ]0,\alpha]$, for all $t\in[t_a,t_b]$, and for all $\varepsilon\in\ ]0,\beta]$.\pesp

The wave function (\ref{W}) gives
\begin{equation}\label{4Ac1}
{\cal S}(x_{\delta\varepsilon}(t),t)=-i\hbar\ln\Psi(x_{\delta\varepsilon}(t),t).
\end{equation}
Let us consider a non relativistic physical system of mass $m$ moving in a time independent potential $\Phi$. The energy is then conserved and the corresponding conservation law gives
\begin{equation}\label{4ener}
E={P^2\over 2m}+\Phi.
\end{equation}
Equations (\ref{4Act2}), (\ref{Cp}) and (\ref{4ener}) imply that the action function ${\cal S}(x_{\delta\varepsilon},t)$ obeys to the Hamilton-Jacobi equation on $[t_a,t_b]$, for $\delta\in\ ]0,\alpha]$, and for all $\varepsilon\in\ ]0,\beta]$
\begin{equation}\label{4HJO}
-{\partial{\cal S}\over\partial t}={1\over 2m}\Big({\partial{\cal S}\over\partial x_{\delta\varepsilon}}\Big)^2+\Phi.
\end{equation}
Direct partial derivatives of the equation (\ref{4Ac1}) give
\begin{equation}\label{4part}
    -{\partial{\cal S}\over\partial t}= i\hbar {\partial\Psi\over\partial t}{1\over\Psi}, \qquad\hbox{and}\qquad {\partial{\cal S}\over\partial x_{\delta\varepsilon} }= -i\hbar {\partial\Psi\over\partial x_{\delta\varepsilon}}{1\over\Psi},
\end{equation}
where the substitution of equations (\ref{4part}) in the Hamilton-Jacobi equation (\ref{4HJO}) gives
\begin{equation}\label{4ln0}
   i\hbar {\partial\Psi\over\partial t}{1\over\Psi}={(-i\hbar)^2\over 2m} \Big({\partial\Psi\over\partial x_{\delta\varepsilon}}\Big)^2{1\over\Psi^2}+\Phi,
\end{equation}
and the partial derivative of the product gives
\begin{equation}
  {\partial \over\partial x_{\delta\varepsilon}}\Big( {\partial\Psi\over\partial x_{\delta\varepsilon}}{1\over\Psi}\Big)={\partial^2\Psi\over\partial x_{\delta\varepsilon}^2}{1\over\Psi}-\Big({\partial\Psi\over\partial x_{\delta\varepsilon}}\Big)^2{1\over\Psi^2},
\end{equation}
that is to say
\begin{equation}\label{4ln1}
 \Big({\partial\Psi\over\partial x_{\delta\varepsilon}}\Big)^2{1\over\Psi^2}={\partial^2\Psi\over\partial x_{\delta\varepsilon}^2}{1\over\Psi}- {\partial \over\partial x_{\delta\varepsilon}}\Big( {\partial\Psi\over\partial x_{\delta\varepsilon}}{1\over\Psi}\Big).
\end{equation}
Thus the substitution of equality (\ref{4ln1}) in the equation (\ref{4ln0}) gives
\begin{equation}\label{4ln2}
   i\hbar {\partial\Psi\over\partial t}{1\over\Psi}={(-i\hbar)^2\over 2m} \Big({\partial^2\Psi\over\partial x_{\delta\varepsilon}^2}{1\over\Psi}- {\partial \over\partial x_{\delta\varepsilon}}\Big( {\partial\Psi\over\partial x_{\delta\varepsilon}}{1\over\Psi}\Big)\Big)+\Phi
\end{equation}
and the multiplication of equation (\ref{4ln2}) by $\Psi$ with a simplification gives
\begin{equation}\label{4ln3}
   i\hbar {\partial\Psi\over\partial t}+{\hbar^2\over 2m}{\partial^2\Psi\over\partial x_{\delta\varepsilon}^2}= \Big(-{\partial \over\partial x_{\delta\varepsilon}}\Big( {\partial\Psi\over\partial x_{\delta\varepsilon}}{1\over\Psi}\Big)+\Phi\Big)\Psi,
\end{equation}
where the use of equations (\ref{4Act1}), (\ref{Cp}), (\ref{4part}), and the chain rule gives
\begin{equation}\label{4term}
-{\partial \over\partial x_{\delta\varepsilon}}\Big( {\partial\Psi\over\partial x_{\delta\varepsilon}}{1\over\Psi}\Big)={-i\over \hbar}{\partial \over \partial t}\Big(m{\cal V}_{\delta\varepsilon}(t)\Big){\partial t\over\partial x_{\delta\varepsilon}}={-im\over \hbar}{\partial {\cal V}_{\delta\varepsilon}(t) \over \partial t}{1\over{\cal V}_{\delta\varepsilon}(t)}.
\end{equation}
The elimination of the term (\ref{4term}) in the right side of the equation (\ref{4ln3}) is possible if one has
\begin{equation}\label{Ass}
{\partial {\cal V}_{\delta\varepsilon}(t) \over \partial t}{1\over{\cal V}_{\delta\varepsilon}(t)}=0
\end{equation}
which is verified since we have a time independent potential $\Phi$, that is to say
\begin{equation}
{\partial^2 {\cal S}\over\partial x_{\delta\varepsilon}^2}=0.
\end{equation}
Then equation (\ref{4ln3}) becomes the Schr\"{o}dinger's equation for a particle moving in a time-independent potential on $[t_a,t_b]$, for $\delta\in\ ]0,\alpha]$, and for all $\varepsilon\in\ ]0,\beta]$
\begin{equation}\label{4ln4}
   i\hbar {\partial\Psi(x_{\delta\varepsilon},t)\over\partial t}+{\hbar^2\over 2m}{\partial^2\Psi(x_{\delta\varepsilon},t)\over\partial x_{\delta\varepsilon}^2}= \Phi\Psi(x_{\delta\varepsilon},t).
\end{equation}

The calculus in this derivation is similar to the one elaborated in \cite{Fie} for a considered wave function $\psi=\exp({iS\over \hbar})$, where the action function is a solution of Hamilton-Jacobi equation. A similar calculus can be traced back to the work elaborated in \cite{CPK} for a considered wave function $\psi=\exp({-iS\over \hbar})$ in which the action function is a solution of the equation ${\partial S\over\partial t}={1\over 2m}\Big({\partial S\over\partial x}\Big)^2 +\Phi$ rather than the Hamilton-Jacobi equation, and the derivation of  the Schr\"odinger's equation was obtained under the same assumption (that is ${\partial^2 S\over\partial x^2}=0$). Several attempts to derive the Schr\"{o}dinger's equation from different principles can be found in the following list but not limited to \cite{HR5},\cite{nel5},\cite{OT5},\cite{PEIC5}.\pesp

The new insight in the above derivation of the Schr\"odinger's equation is the use of an infinity of paths connected to each other via the geometry of a given space-time print, and the assumption of considering the time independent potential $\Phi$ means, using (\ref{Vde}) and (\ref{Ass}), that on $[t_a,t_b]$, for $\delta\in\ ]0,\alpha]$, and for all $\varepsilon\in\ ]0,\beta]$
\begin{equation}\label{Vk}
{\partial {\cal V}_{\delta\varepsilon}(t) \over \partial t}{1\over{\cal V}_{\delta\varepsilon}(t)}=0,
\end{equation}
that is to say for $\delta\in\ ]0,\alpha]$, for all $t\in[t_a,t_b]$, and for all $\varepsilon\in\ ]0,\beta]$
\begin{equation}\label{nsc}
\ln (\vert{\cal V}_{\delta\varepsilon}(t)\vert)=cst,
\end{equation}
which is the necessary and sufficient condition to derive the Schr\"{o}dinger's equation for a non-relativistic physical system moving in a time independent potential starling from Jacobi-Hamilton equation.
\pesp
Each probability amplitude $\Psi(x_{\delta\varepsilon},t)$ associated to each trajectory $x_{\delta\varepsilon}(t)$ is found to satisfy the Schr\"{o}dinger's equation for a non-relativistic physical system moving in a time independent potential.

\section{Prospect}

 The different steps of this approach to the Schr\"{o}dinger's formulation of quantum mechanics for a non relativistic physical system moving in time independent potential reside in the following: \pesp

(i) Construction of all possible paths between two distant locations that follow the geometry of the space-time print characterized by a geodesics defined by a continuous function (this function can be a continuous and nowhere differentiable function). \pesp

(ii) Association of a wave function (\ref{W}) as a probability amplitude for each path identical to the Feynman's probability amplitude via a functional action subject to the equations (\ref{4Act1}) and (\ref{4Act2}), compatible with the correspondence principles of quantum mechanics for momentum and total energy, in which each path is represented by the same amplitude, meanwhile the different paths depend on the classical action compatible with the Feynman's postulate II.\pesp

(iii) Each associated probability amplitude (\ref{W}) for each path is found to be compatible with the Schr\"{o}dinger's equation (\ref{4ln4}), for a non relativistic physical system moving in a time independent potential, starting from the Jacobi-Hamilton equation.\pesp

This approach is based on a formalism introduced in the recent research work \cite{BFP} where the set of continuous and nowhere differentiable functions has been presented as a candidate for the quantum world. This work comforts that assertion since this approach remains valid for a given space-time print with more constraints such as fluctuation, points of non differentiability, or a nowhere differentiability.

\bibliographystyle{spmpsci}
\bibliography{allbiblio}

\end{document}